# Long-Term Mobile Traffic Forecasting Using Deep Spatio-Temporal Neural Networks

Chaoyun Zhang
School of Informatics
chaoyun.zhang@ed.ac.uk

Paul Patras
School of Informatics
paul.patras@ed.ac.uk

## ABSTRACT

Forecasting with high accuracy the volume of data traffic that mobile users will consume is becoming increasingly important for precision traffic engineering, demand-aware network resource allocation, as well as public transportation. Measurements collection in dense urban deployments is however complex and expensive, and the post-processing required to make predictions is highly non-trivial, given the intricate spatio-temporal variability of mobile traffic due to user mobility. To overcome these challenges, in this paper we harness the exceptional feature extraction abilities of deep learning and propose a Spatio-Temporal neural Network (STN) architecture purposely designed for *precise network-wide mobile traffic forecasting*. We present a mechanism that fine tunes the STN and enables its operation with only limited ground truth observations. We then introduce a Double STN technique (D-STN), which uniquely combines the STN predictions with historical statistics, thereby making faithful *long-term* mobile traffic projections. Experiments we conduct with real-world mobile traffic data sets, collected over 60 days in both urban and rural areas, demonstrate that the proposed (D-)STN schemes perform up to 10-hour long predictions with remarkable accuracy, irrespective of the time of day when they are triggered. Specifically, our solutions achieve up to 61% smaller prediction errors as compared to widely used forecasting approaches, while operating with up to 600 times shorter measurement intervals.

## KEYWORDS

mobile traffic forecasting, deep learning, spatio-temporal modelling

## 1 INTRODUCTION

The annual mobile traffic consumption will exceed half a zettabyte by 2021, an almost 5-fold increase of the current demand [4]. Emerging applications including augmented/virtual reality, autonomous vehicles, and digital healthcare will account for a significant fraction of future demands. To meet the stringent performance requirements of such applications, precision traffic engineering and demand-aware allocation of cellular network resources become essential. These tasks require real-time traffic analysis and accurate prediction capabilities [33], which are challenging to implement with existing tools [21]. In particular, mobile network monitoring currently relies on specialised equipment, e.g. probes [6, 14]. Deploying these at each base station is expensive and involves storing locally massive amounts of logs that later have to be transfered for analysis. If monitoring is instead exclusively employed at selected locations in the core network, it requires substantial processing power. Timely and exact mobile traffic forecasting is further complicated by the complex spatio-temporal patterns of user demand [8, 29].

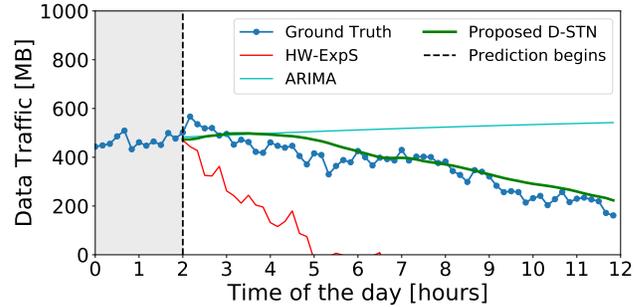

Figure 1: Traffic volume in one cell forecast over 10 hours (after 2-hour observations) with Holt-Winters Exponential Smoothing (HW-ExpS), ARIMA, and the original deep learning based approach we introduce in this paper (D-STN), which captures spatio-temporal correlations. Experiments with the Telecom Italia data set for Milan [2].

Alarm based monitoring systems enable engineers to react to abrupt changes in traffic volume only *a posteriori*, which impacts negatively on the latency perceived by interactive applications. While long-term network traffic forecasting techniques have been proposed to overcome this problem in wired broadband networks (e.g. [17, 22]), mobile networks have received less attention [20]. Existing mobile traffic prediction mechanisms (e.g. [15, 27]) work well in estimating *trends*, though largely undertake naïve predictive modelling across time series observed at *individual* base stations; therefore, they are ill-suited to network-wide forecasting. Further, they ignore important spatial correlations associated with user movement, hence are predominantly limited to short-term predictions. Such is the case of common practice exponential smoothing (ExpS) [27] and Auto-Regressive Integrated Moving Average (ARIMA) [15] techniques, whose accuracy degrades considerably over time when the data series exhibit frequent fluctuations, as we exemplify in Fig. 1. In the figure we also illustrate the behaviour of an original deep learning based approach that we introduce in this paper. Our proposal exploits cross-spatial and cross-temporal features of the city-wide traffic consumption, and operates substantially longer in the absence of timely measurements, significantly outperforming prior schemes in terms of prediction accuracy.

**The Potential of Deep Learning:** Underpinned by recent advances in GPU computing and stochastic optimisation algorithmics [10, 16], machine learning and neural network architectures achieve remarkable results in image recognition [18] natural language processing [5] and mobile traffic analytics [34]. Structures such as the Long Short-Term Memory (LSTM) [9] work well in applications that operate with sequential data, including speech recognition [11] and machine translation [26]. By adding a convolutional layer to LSTMs, the ConvLSTM enables capturing spatial relationships between data points for precipitation nowcasting [32]. Spatio-temporal information is also exploited by 3D-ConvNets [13], which add a third dimension to conventional convolutional neural networks to perform video recognition. We observe important similarities between abstract representations of moving images and sequences of mobile traffic snapshots in different coverage areas. Thus the potential of *deep learning for reliable long-term mobile data traffic forecasting* awaits to be explored.

**Contributions:** We propose an original deep Spatio-Temporal neural Network (STN) architecture that exploits important correlations between user traffic patterns at different locations and times, to achieve *precise network-wide mobile traffic forecasting* and overcome the limitations of commonly used prediction techniques. Specifically, we harness the ability of ConvLSTM and 3D-ConvNet structures to model long-term trends and short-term variations of the mobile traffic volume, respectively. We build an ensembling system that exploits the benefits of both models by using two layers that fuse the features extracted by each model from previous traffic measurements. We then employ a multilayer perceptron (MLP) [10] to map the output of the second fusion layer onto final predictions of future mobile traffic volumes. To the best of our knowledge, this is the first time such neural network structures are fused and employed for the purpose of mobile traffic forecasting. The STN is demonstrably effective in spatio-temporal features extraction through convolution operations, while being less complex than traditional neural networks such Restricted Boltzmann Machines (RBM), which require a considerably larger number of parameters to be stored in memory [25].

Secondly, to enable *long-term traffic forecasting with only limited observations*, we propose an Ouroboros Training Scheme (OTS). This fine tunes the deep neural network such that earlier predictions can be used as input, while the difference between its output and the ground truth is minimised. Our intuition is that re-training the neural network on actual traffic measurements combined with one-step predictions made over these will enhances long-term predictions' quality when ground truth becomes (partly) unavailable.

Thirdly, we specify a Double STN (D-STN) solution that combines the proposed STN with a decay mechanism, which mitigates accumulating prediction errors by mixing the predictions with an empirical mean of the locally observed traffic. This approach ensures mobile traffic volume forecasts remain within reasonable bounds and their duration is substantially extended. Unlike recent work that combines LSTMs and autoencoders for traffic prediction purposes [30], the proposed (D-)STN are not limited to 1-step inferences. Instead, we achieve practical and reliable multi-step forecasting without requiring to train separate neural networks for each base station.

Finally, we implement the proposed (D-)STN prediction techniques on a GPU cluster and conduct experiments on publicly available real-world mobile traffic data sets collected over 60 days and released through the Telecom Italia's Big Data Challenge [2]. The results obtained demonstrate that, once trained, our solutions provide high-accuracy long-term (10-hour long) traffic predictions, while operating with short observation intervals (2 hours) and irrespective of the time of day when they are triggered. Importantly, our models outperform commonly used prediction methods (HW-ExpS and ARIMA), as well as the traditional multilayer perceptron (MLP) and Support Vector Machine (SVM), reducing the normalised root mean square error (NRMSE) by up to 61% and requiring up to 600 times shorter measurement intervals.

## 2 THE MOBILE TRAFFIC FORECASTING PROBLEM

Our objective is to make accurate long-term forecasts of the volume of mobile data traffic users consume at different locations in a city, following measurement-based observations. We formally express network-wide mobile traffic consumption observed over a time interval $T$ as a spatio-temporal sequence of data points $\mathcal{D} = \{D_1, D_2, ..., D_T\}$, where $D_t$ is a snapshot at time $t$ of the mobile traffic volume in a geographical region represented as an $X \times Y$ grid, i.e.

$$D_t = \begin{bmatrix} d_t^{(1,1)} & \cdots & d_t^{(1,Y)} \\ \vdots & \vdots & \vdots \\ d_t^{(X,1)} & \cdots & d_t^{(X,Y)} \end{bmatrix}, \quad (1)$$

where $d_t^{(x,y)}$ measures the data traffic volume in a square cell with coordinates $(x, y)$ and the sequence can be regarded as a tensor $\mathcal{D} \in \mathbb{R}^{T \times X \times Y}$. From a machine learning perspective, the spatio-temporal traffic forecasting problem is to predict the most likely $K$-step sequence of data points, given previous $S$ observations. That means solving

$$\widetilde{D}_{t+1}, \ldots, \widetilde{D}_{t+K} = \underset{D_{t+1}, \ldots, D_{t+K}}{\arg\max} \; p(D_{t+1}, \ldots, D_{t+K} | D_{t-S+1}, \ldots, D_t). \quad (2)$$

There is growing evidence that important spatio-temporal correlations exist between traffic patterns [8, 29], though the value at any location, $d_{t+1}^{(x,y)}$, largely depends only on the traffic in neighbouring cells and information associated with distant cells could be neglected [20]. That is, statistical dependence exists between proximate cells, while the traffic patterns at distant cells provide little insight into how traffic consumption will evolve at a 'target' cell. Therefore, confining consideration to the traffic in $(r+1) \times (r+1)$ adjacent cells allows us to simplify the problem and express one-step predictions as

$$p(D_{t+1} | D_{t-S+1}, \ldots, D_t) \approx \prod_{x=1}^{X} \prod_{y=1}^{Y} p\left(d_{t+1}^{(x,y)} | F_{t-S+1}^{(x,y)}, \ldots, F_t^{(x,y)}\right), \quad (3)$$

where

$$F_t^{(x,y)} = \begin{bmatrix} d_t^{(x-\frac{r}{2}, y-\frac{r}{2})} & \cdots & d_t^{(x+\frac{r}{2}, y-\frac{r}{2})} \\ \vdots & d_t^{(x,y)} & \vdots \\ d_t^{(x-\frac{r}{2}, y+\frac{r}{2})} & \cdots & d_t^{(x+\frac{r}{2}, y+\frac{r}{2})} \end{bmatrix} \quad (4)$$

is the data traffic matrix at time $t$ in an $(r+1) \times (r+1)$ region adjacent to location $(x, y)$. Then the prediction of $D_{t+1}$ can be expressed as the set

$$\widetilde{D}_{t+1} = \left\{ \widetilde{d}_{t+1}^{(x,y)} \mid x = 1, \ldots, X; y = 1, \ldots, Y \right\}, \quad (5)$$

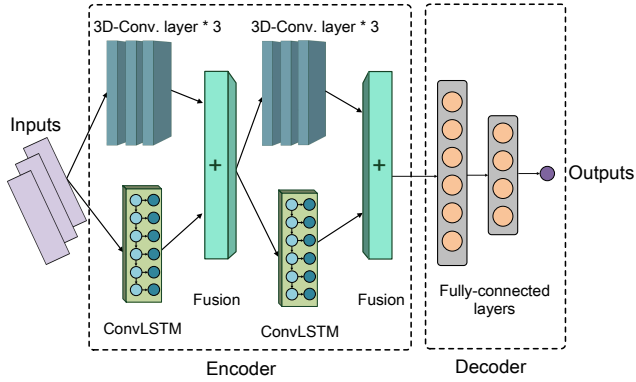

**Figure 2: The encoder-decoder architecture of the proposed STN. Measured traffic matrices are processed by ConvLSTMs and 3D-ConvNets to encode spatio-temporal features, then fused. Subsequently, the intermediary output is decoded into predictions by a stack of fully-connected layers (MLP).**

where $\widetilde{d}_{t+1}^{(x,y)}$ is the prediction for $d_{t+1}^{(x,y)}$ and is obtained by solving

$$\widetilde{d}_{t+1}^{(x,y)} = \arg\max_{d_{t+1}^{(x,y)}} p\left(d_{t+1}^{(x,y)} | F_{t-S+1}^{(x,y)}, \ldots, F_t^{(x,y)}\right). \quad (6)$$

The above predictions depend only on a marginal distribution, which is precisely the distribution we will model with the Spatio-Temporal Network (STN) we propose next to forecast mobile traffic across different locations and times.

## 3 THE SPATIO-TEMPORAL NETWORK

We design a deep neural network architecture, which we name Spatio-Temporal neural Network (STN), to solve the traffic forecasting problem posed in Sec. 2. The proposed STN follows an encoder-decoder paradigm, where we combine a stack of Convolutional Long Short-Term Memory (ConvLSTM) and three-dimensional Convolutional Network (3D-ConvNet) elements, as illustrated in Fig. 2 and detailed next. Our intuition is that the ability of these structures to handle time series data with spatial dependencies as already demonstrated in e.g. video applications, could be exploited for accurate mobile traffic forecasting. In our case each of these elements are fed with traffic matrices (4) and embed this input through hidden layers into several feature maps, which we then fuse and return at the output of the encoder. The decoder is a multi-layer perceptron (MLP), which is a supervised learning technique that takes as input the output of the encoder and makes the final predictions through fully-connected layers, as specified in (6). The key benefit of employing the MLP lies within the model's ability to solve complex regression problems. To our knowledge, *the problem of precise mobile traffic forecasting has not been tackled previously by fusing ConvLSTM and 3D-ConvNet neural networks*, as we propose. In what follows we outline the operation of these structures and explain how they are fused.

**ConvLSTMs:** The Long Short-Term Memory (LSTM) is a special recurrent neural network (RNN) that remedies the vanishing gradient problem characteristic to RNNs, by introducing a set of "gates" [9]. The gates in standard LSTMs are usually fully connected and as a consequence the number of parameters is large, requiring substantial memory and computation time for training purposes. Hence, although proven powerful when working with sequential data, this model is highly complex and frequently turns overfitted.

ConvLSTMs solve this problem by replacing the inner dense connections with convolution operations [32]. This approach reduces significantly the number of parameters in the model and enhances its ability to handle spatio-temporal data, which is particularly important to our problem. Further, ConvLSTMs can capture long-term trends present in sequences of data points, which makes them particularly well-suited to making inferences about mobile data traffic, as it known this exhibits important spatio-temporal correlations [29, 33].

Given a sequence of 3-D inputs (i.e. multiple "frames" of data traffic measurements across 2-D grids/feature maps) denoted $\mathbf{X} = \{X_1, X_2, \ldots, X_T\}$, we specify the operations of a single ConvLSTM in (7). Here '$\odot$' denotes the Hadamard product, '$*$' the 2-D convolution operator, and $\sigma(\cdot)$ is a sigmoid function. Since each hidden element of this neural network is represented as a 2-D map, we effectively capture cross-spatial traffic correlations through the convolution operations.

$$\begin{aligned} i_t &= \sigma(W_{xi} * X_t + W_{hi} * H_{t-1} + W_{ci} \odot C_{t-1} + b_i), \\ f_t &= \sigma(W_{xf} * X_t + W_{hf} * H_{t-1} + W_{cf} \odot C_{t-1} + b_f), \\ C_t &= f_t \odot C_{t-1} + i_t \odot \tanh(W_{xc} * X_t + W_{hc} * H_{t-1} + b_c), \quad (7) \\ o_t &= \sigma(W_{xo} * X_t + W_{ho} * H_{t-1} + W_{co} \odot C_t + b_o), \\ H_t &= o_t \odot \tanh(C_t). \end{aligned}$$

In the above, $W_{(\cdot\cdot)}$ and $b_{(\cdot)}$ denote weights and biases we obtain through model training, which we perform on the complete architecture using a stochastic optimisation algorithm, as we detail towards the end of this section. Note that the inputs $X_t$, cell outputs $C_t$, hidden states $H_t$, input gates $i_t$, forget gates $f_t$, and output gates $o_t$ in the ConvLSTM's inner structure are all 3-D tensors. The first two dimensions of the tensors form the spatial dimension, while the third is the number of feature maps. The input-to-state, cell-to-state, and cell-to-cell transitions are element-wise controlled by each gate $i_t, o_t$, and $f_t$, which allows the model to "learn to forget" in the spatio-temporal dimension. This property dramatically improves the model's ability to capture spatio-temporal trends [32].

**3D-ConvNets:** Our STN architecture further includes 3D-ConvNet elements (see Fig. 2), which extend standard ConvNet models with a temporal dimension [13]. This choice is motivated by recent results showing 3D-ConvNet perform remarkably in terms of spatio-temporal feature learning [12, 13]. Further, they also capture well local dependencies, as seen in minor fluctuations of traffic sequences that are triggered by stochastic human mobility. Given a sequence of spatio-temporal data with $N$ feature maps $\mathbf{X} = \{X_1, X_2, \ldots, X_N\}$, the output of a 3-D convolutional layer will consist of $H_1, \ldots H_M$ convoluted feature maps, given by

$$H_m = \text{act}\left(\sum_{n=1}^{N} X_n * W_{mn} + b_m\right), \quad (8)$$

where '$*$' corresponds now to a 3-D convolution operator and act($\cdot$) denotes an activation function, whose objective is to increase the model's non-linearity. Functions such as the rectified linear unit (ReLU) and the sigmoid are commonly used for this purpose. We note that 3D-ConvNets differ from ConvLSTMs primarily because they do not involve back-propagation through time (BPTT), which in the latter happens between cells. Instead 3D-ConvNets require more layers to attain similar results. On the other hand, the 3-D convolutions enable the model to also capture cross-temporal traffic correlations essential in our problem. The 3D-ConvNet shares weights across different locations in the input, allowing to maintain the relation between neighbouring input points and spatio-temporal locality in feature representations. This property enables 3D-ConvNets to capture better short-term traffic fluctuations and overall improves the model's generalisation abilities.

**STN – Fusing ConvLSTMs and 3D-ConvNets:** To leverage the ability of both ConvLSTM and 3D-ConvNet to learn spatio-temporal features and forecast mobile data traffic with high accuracy, the STN architecture we propose blends the output of both models through two "fusion" layers (see Fig. 2). The goal of the fusion operation is to build an ensembling system which includes two dedicated deep spatio-temporal models, which has been proven to enhance the model's performance [28]. By averaging intermediate outputs of both ConvLSTM and 3D-ConvNet twice, the proposed STN reinforces the ensembling system, allowing it to integrate the advantages of both architectures, thereby improving the robustness of the model. Through the fusion operations we jointly exploit the advantages of both ConvLSTM (capturing long-term trends) and 3D-ConvNet (capturing local fluctuations), which leads to superior prediction performance, as compared to simply employing any of the two individually. This is demonstrated by the results we present in Sec. 5 and is a key novelty of our purpose-built neural network architecture. Importantly, unlike other deep learning approaches proposed recently [30], where different structures extract spatial and temporal patterns separately, our solution can jointly distil spatio-temporal traffic features. This enables us to train our architecture in an end-to-end manner, instead of training each of its component individually, as required in [30]. Further, the proposed STN has excellent generalisation abilities, as we will demonstrate it is sufficient to train for a single geographic area, before applying it for inference to other.

In our design the input data (measurements) is first processed in parallel by one ConvLSTM and one 3D-ConvNet, then their outputs are aggregated by a fusion layer that performs the following element-wise addition:

$$H(\Theta_H; X) = h_C(\Theta_1; X) + h_L(\Theta_2; X), \quad (9)$$

where $h_C$ and $h_L$ are the outputs of a single 3D-ConvNet and respectively a ConvLSTM, and $\Theta_H = \{\Theta_1, \Theta_2\}$ denotes the set of their parameters (weights and biases). We employ this procedure twice to encode the spatio-temporal embedding of the data sequence given as input. Subsequently, we decode the obtained features and perform prediction via an MLP. The MLP connects the outputs from the second fusion layer over every time step to achieve an "attention"-like mechanism [26]. This allows the model to make use of all temporal features learned to produce final predictions, rather than merely relying on the last state. It is important to note that, unlike traditional Restricted Boltzmann Machines (RBMs) employed for time series forecasting [19], the proposed STN shares weights between inputs, thus requiring to configure considerably fewer parameters.

The STN models the expected value of the marginal distribution $p(d_{t+1}^{(x,y)}|F_{t-S+1}^{(x,y)}, \ldots, F_t^{(x,y)})$, i.e. it takes a "local" spatio-temporal traffic matrix to predict the volume of traffic at the following time instance $t + 1$, and its output consists only of one regression value corresponding to an $(x, y)$ location. The prediction is repeated multiple times to encompass all the locations covered by the cellular network grid.

The encoder contains six 3D-convolutional layers with (3, 3, 3) and respectively (6, 6, 6) feature maps, and two ConvLSTM layers with 3 and 6 feature maps. In our design, we set the length of the input $S = 12$ (corresponding to 2 hours, with traffic volume snapshots every 10 mins) and $r = 10$ (i.e. considering spatial correlation among 10×10 adjacent cells). Thus each input is a $11 \times 11 \times 12$ tensor and the STN model will predict the traffic volume $\widetilde{d}_{t+1}$ at the centre of a $11 \times 11$ map in the next time step.

To obtain these predictions, we effectively train a neural network model $\mathcal{M}$ parameterised by $\Theta$, which takes a set of inputs $\mathbf{F}_t^{(x,y)} := \left\{F_{t-S+1}^{(x,y)}, F_{t-S+2}^{(x,y)}, \ldots, F_t^{(x,y)}\right\}$ and outputs traffic volume forecasts $\widetilde{d}_{t+1}^{(x,y)}$, i.e.

$$\widetilde{d}_{t+1}^{(x,y)} = \mathcal{M}\left(\Theta; \mathbf{F}_t^{(x,y)}\right). \quad (10)$$

We adopt the maximum likelihood estimation (MLE) method to train this model by minimising the following $L_2$ loss function:

$$L(\Theta) = \frac{1}{T \cdot X \cdot Y} \sum_{t=1}^{T} \sum_{x=1}^{X} \sum_{y=1}^{Y} ||\mathcal{M}(\Theta; \mathbf{F}_t^{(x,y)}) - d_t^{(x,y)}||^2. \quad (11)$$

Stochastic Gradient Descent (SGD) based methods are widely used to optimise this loss function. In our work, we choose the Adam optimiser [16], which commonly yields faster convergence compared to traditional SGD. We employ standard configuration ($\beta_1 = 0.9, \beta_2 = 0.999, \epsilon = 10^{-8}$), setting the initial learning rate to 0.005.

Note that STN only performs *one-step* predictions based on complete ground truth information available at every step. This is however problematic when multi-step prediction is desired, as soon as ground truth information becomes unavailable. In the next section we extend our design to achieve multi-step predictions and *forecast mobile traffic long-term with only limited ground truth observations*.

## 4 LONG-TERM MOBILE TRAFFIC FORECASTING

In the case of multi-step predictions, ground truth becomes rapidly unavailable, while still necessary for conditioning in the prediction problem posed in (6). An intuitive way to address this issue is to recursively reuse recent predictions as input for the following prediction step. Inevitably, this leads to prediction errors that are likely to accumulate over time and the results obtained may be modest. For instance, the predictions the proposed STN makes will be very accurate when ground truth measurements are always available, as we illustrate in the shaded region of Fig. 3. However, when network measurements are suspended, the pure STN's output is rapidly unable to follow the actual evolution of the data traffic (observe the red curve in the light region shown in Fig. 3).

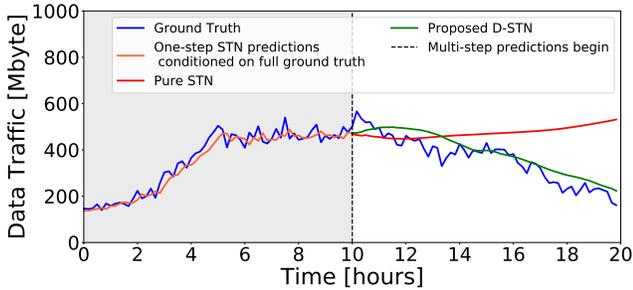

Figure 3: Traffic volume predicted by pure STN and D-STN. First (shaded area), the STN makes one-step predictions based on fully observable ground truth, then both methods perform multi-step predictions as the ground truth becomes unavailable.

To address this problem and achieve reliable long-term forecasting, we propose a Double STN scheme (D-STN), which comprises two key enhancements. Specifically, (1) we introduce an Ouroboros Training Scheme (OTS) that fine tunes the neural network, allowing earlier predictions to be fed as inputs when ground truth becomes unavailable, while keeping prediction errors small; (2) D-STN blends the newly trained STN with historical statistics that essentially summarise prior knowledge of long-term trends, and thus enables accurate forecasting for extended periods of time (up to 10h). We exemplify the D-STN's ability to track the real traffic in Fig. 3 and detail its operation next.

## 4.1 An Ouroboros Training Scheme

Turning attention again to (6), we can see that ground truth observations become unavailable as early as at the second prediction step. If working with $S = 12$, by the twelfth step, no measurements will be available for prediction. To deal with this issue, one possibility is to substitute the missing information with earlier model predictions. However, simply conditioning on inferred data will unavoidably introduce errors that accumulate over time. Intuitively, the reason is that the model only works well when the training data and test data follow the same distributions. When the model operates based on its own predictions, any prediction error, however small, will lead to diverging distributions of input and respectively real data. Similar behaviour was previously observed in imitation learning problems [23]. Examining again Fig. 3, observe the traffic volume experiences an abrupt rise when the multi-step prediction is activated. As a result, the pure STN, which completely relies on the model's predictions, fails to capture this initial trend and ultimately the error grows considerably with the prediction step.

We solve this problem by introducing an Ouroboros training scheme (OTS), which draws inspiration from the DAGGER technique that mixes ground truth and self-generated sequences in the model training phase [24]. Our goal is to enable the neural network to forecast precisely, irrespective of whether conditioning on ground truth or on predictions it made earlier. Therefore, apart from the original training data, the OTS recursively stores the model predictions and uses these for a second round of training, to *mimic the prediction behaviour*. We summarise this procedure in Algorithm 1, where at every epoch $e$, we regard the inputs fed to the model as a queue $Q$. First ($t = 1$) we initialise this with $S$ ground truth observations (line 6). Subsequently, at every sub-step $t$, we pop the oldest frame in $Q$ (line 8) and push the prediction made at the last step (lines 9–10), thereby rebuilding the model's input. When used, $Q$ is partitioned into $X \times Y$ data points, each of which is an $(r+1) \times (r+1) \times S$ tensor (in our design these are 10,000 tensors with dimension $11 \times 11 \times 12$). We feed the model with $Q$ and corresponding ground truth $T$, then preform Stochastic Gradient Descent (SGD) based training [16] (line 13). An epoch will stop when it exhausts all training data, i.e. $t$ reaches $T - S$. This procedure resembles the behaviour of a mythological snake called Ouroboros, which perpetually eats its own tail (hence the name of the proposed scheme). We work with $E = 1$ epoch, which is sufficient for our problem.

The OTS algorithm works particularly well, as it broadens the horizon of the model by extending the training data with its predictions and enlarging the support set of the input distribution. This can be regarded as a data augmentation technique, which was proven powerful when training complex models [18]. In our case OTS suppresses the overestimation tendency of the pure STN, as the retraining forces the predictions to be substantially closer to the ground truth.

## 4.2 Blending Predictions and Historical Statistics

Employing the OTS will improve the accuracy of multi-step predictions, though note that uncertainty may grow over time and thus limit accuracy when long-term (e.g. >5h long) forecasting is desired. In what follows we propose a further improved Double STN (D-STN) forecasting system that addresses this issue.

The D-STN design stems from two important properties of mobile data traffic we observed. Namely, data traffic exhibits certain

---

**Algorithm 1** The Ouroboros Training Scheme

1: **Inputs:**
   Time series training data $\mathcal{D} = \{D_1, D_2, ..., D_T\}$
2: **Initialize:**
   A pre-trained model $\mathcal{M}$ with parameters $\Theta$.
3: **for** $e = 1$ to $E$ **do**
4:   **for** $t = 1$ to $T - S$ **do**
5:     **if** $t = 1$ **then**
6:       $Q \leftarrow \{D_1, D_2, ..., D_S\}$ ▶ Generate input queue using first $S$ ground truth measurements.
7:     **else**
8:       Pop the first element out of $Q$,
9:       Predict $D'_{S+t-1}$ by $\mathcal{M}$ with input $Q$,
10:      Push $D'_{S+t-1}$ to the end of $Q$.
11:     **end if**
12:     Generate the target input $T \leftarrow D_{S+t}$.
13:     Train $\mathcal{M}$ with input $Q$ and target $T$ by SGD.
14:   **end for**
15: **end for**

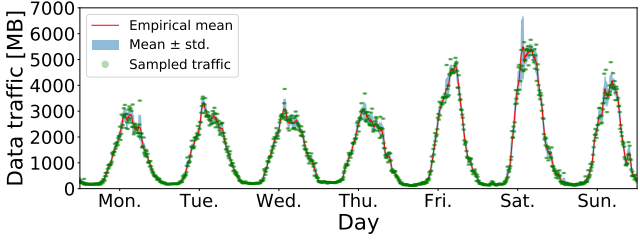

Figure 4: Traffic volume sampled over 7 weeks (1$^{st}$ Nov – 12$^{th}$ Dec 2013), weekly empirical mean, and standard deviation in cell (48, 60) of the Milan grid.

periodicity (in both daily and weekly patterns) and relatively flat averages, if observed over long intervals [20]. This is indeed the case also for the city of Milan, as we illustrate in Fig. 4, where we plot the weekly empirical mean of mobile traffic volume in a selected cell between 1 Nov and 12 Dec 2013 (7 weeks). The figure also shows the actual sampled traffic and the standard deviation in the same period. Observe that despite a few outliers, traffic volume samples are close to the empirical mean. We conjecture that incorporating prior knowledge of averages into the model can reduce uncertainty and improve the prediction performance, if utilised appropriately. On the other hand, there is little correlation between traffic volumes measured at intervals far (e.g. hours) apart, which has also been observed for other similar time series [3].

Therefore, we propose to combine information of the weekly empirical mean with the STN predictions, using a light-weight decay mechanism that reduces the weight of the predictions over time. In our experiments we use the weekly empirical mean of mobile traffic volume computed over 7 weeks (1 Nov – 12 Dec 2013). Assume we trigger forecasting at time $t$ and denote $D^M_{t+h}$ and $D^O_{t+h}$ the predictions at time $t + h$ of the original STN without fine tuning and respectively with the OTS enabled. Denoting $\overline{D}_{t+h}$ the weekly empirical mean of actual measurements, we define the $h$-step ahead forecasting of our D-STN as:

$$D'_{t+h} := \gamma(h) \left[ \alpha(h) D^M_{t+h} + (1 - \alpha(h)) D^O_{t+h} \right] + (1 - \gamma(h)) \overline{D}_{t+h}, \quad (12)$$

where

$$\gamma(h) = 1 - \frac{1}{1 + e^{-(wh+b)}} \quad (13)$$

is a sigmoid decay function which is rectified between (0, 1). $\gamma(h)$ controls the weighting of model predictions and empirical means. It will non-linearly degenerate to zero over $h$, meaning the empirical mean will dominate the prediction's result if $h$ is large. The hyperparameters $w$ and $b$ control the changing rate and the initial state of $\gamma(h)$, and are set to $w = 0.01$ and $b = -5$ based on **cross validation over training set**. Note that $b$ must be a large negative value to guarantee that the model's predictions will mostly contribute at the beginning of the process. In practice, the empirical mean $\overline{D}_{t+h}$ can also be updated in an on-line manner, as new measurements are conducted.

In addition, the predictions made by the two models $D^M_{t+h}$ and $D^O_{t+h}$ are weighted by $\alpha(h)$, which is also dynamically updated as follows:

$$\alpha(h) = \max\left(1 - \frac{h(1-\delta)}{S}, \delta\right). \quad (14)$$

Recall that $S$ denotes the temporal length of the input sequence and $\delta$ is a fixed lower bound for $\alpha$. Since the fine tuned STN is trained using OTS with the model predictions, the original STN deserves a heavier weight in the initial steps, where the model's input still consists partially of ground truth observations. To ensure $D^M_{t+h}$ and $D^O_{t+h}$ contribute equally as time advances, we set $\delta = 0.5$. The weights of $D^M_{t+h}, D^O_{t+h}$ and $\overline{D}_{t+h}$, sum up to 1, i.e.

$$\gamma(h) \times \alpha(h) + \gamma(h) \times (1 - \alpha(h)) + 1 - \gamma(h) = 1, \quad (15)$$

which normalises the output of the D-STN. Note that the original and the OTS trained models share the same input when making predictions, i.e.

$$\mathcal{D} = \begin{cases} \{D_{t+h-S}, ..., D_t\} \cup \{D'_{t+1}, ..., D'_{t+h-1}\}, & h \leqslant S; \\ \{D'_{t+h-S}, ..., D'_{t+h-1}\}, & h > S, \end{cases} \quad (16)$$

where $D_t$ is the observable ground truth at time $t$.

Next we compare the performance of STN, D-STN, commonly used network traffic prediction schemes, and other deep learning based predictors, demonstrating substantial accuracy gains over state-of-the-art techniques.

## 5 PERFORMANCE EVALUATION

In this section we first briefly describe our implementation of the proposed neural network models. Then we evaluate the performance of STN and D-STN by conducting multi-step predictions and comparing their accuracy with that of Holt-Winters Exponential Smoothing (HW-ExpS), Auto-Regressive Integrated Moving Average (ARIMA), the Multi-Layer Perceptron (MLP), standard ConvLSTM and 3D-ConvNet based predictors, and recent deep learning-based short-term predictors.

### 5.1 Mobile Traffic Data Sets

We experiment with publicly available real-world mobile traffic data sets released through Telecom Italia's Big Data Challenge [2]. These contain network activity measurements in terms of total cellular traffic volume observed over 10-minute intervals, for the city of Milan and the Trentino region, collected between 1 Nov 2013 and 1 Jan 2014 (2 months). The two areas are of different sizes, have different population, and therefore exhibit dissimilar traffic patterns. Milan's coverage area is partitioned into $100 \times 100$ squares of 0.055km$^2$ (i.e. 235m × 235m). Trentino's coverage is composed of $117 \times 98$ cells of 1km$^2$ each.

### 5.2 Implementation

We implement the proposed (D-)STN forecasting schemes as well as the conventional ConvLSTM, 3D-ConvNet and MLP using the open-source Python libraries TensorFlow [1] and TensorLayer [7]. We train the models using a GPU cluster comprising 15 nodes, each equipped with 1-2 NVIDIA TITAN X and Tesla K40M computing accelerators (3584 and respectively 2280 cores). To evaluate their prediction performance, we use one machine with one NVIDIA GeForce GTX 970M GPU for acceleration (1280 cores).

We implement, train, and evaluate the ARIMA model using the `statsmodels` Python package without GPU acceleration. Training the HW-ExpS technique is fairly straightforward, as the procedure involves a simple loop to compute the sequence of smoothed values, best estimates of the linear trend, and seasonal correction factors [31].

## 5.3 Prediction Accuracy Evaluation and Comparison with Existing Techniques

Next we evaluate the performance of our proposals against widely used time series prediction techniques, namely HW-ExpS configured with $\alpha = 0.9, \beta = 0.1$, and $\gamma = 0.001$, ARIMA using $p = 3, d = 1$, and $q = 2$, and respectively MLP with 2 layers. The hyper-parameters of HW-ExpS and ARIMA are selected based on **cross validation on training sets**. To achieve a fair comparison, we stop feeding HW-ExpS and ARIMA with ground truth for further tuning when prediction is triggered. We also evaluate the performances of two components of the STN, namely the traditional ConvLSTM and 3D-ConvNet. Arguably other traditional machine learning tools such as Deep Belief Networks (DBNs – stacks of Restricted Boltzmann Machines) and Support Vector Machine (SVMs) could also be employed for comparison. However, we give limited consideration to these models, since DBNs are essentially MLPs that perform layer-wise pre-training to initialise the weights, which requires substantially more time for the benefit of reducing overfitting (this is not viewed as essential, especially when the data used for training is sufficient). The standard SVM training algorithm has $O(n^3)$ and $O(n^2)$ time and space complexities, where $n$ is the size of the training set (in our case, $n = 57,480,000$ points), and is thus computationally infeasible on very large data sets. We do however compare the prediction performance of our schemes against that of an SVM trained on a sub-set of the original data. For completion, we also compare our proposal against a recent deep learning approach [30], which comprises auto-encoders (AEs) and an LSTM. Unfortunately, this solution cannot operate when ground truth data is (partially) unavailable and is only able to perform one-step predictions. Therefore in this case we limit the comparison to such short-term forecasts.

We train all deep learning models, i.e. (D-)STN, ConvLSTM, 3D-ConvNet, MLP, and AE+LSTM with data collected in Milan between 1 Nov 2013 and 10 Dec 2013 (40 days), validate on the data in the same city for the following 10 days, and we evaluate their performances on both Milan and Trentino data sets collected between 20–30 Dec 2013. We train HW-ExpS and ARIMA on both data sets with measurements collected during the first 50 days and test on the same sets used for evaluating the deep learning models.

We quantify the accuracy of the proposed (D-)STN and existing prediction methods through the Normalised Root Mean Square Error (NRMSE) as given below:

$$\text{NRMSE} = \frac{1}{\bar{d}}\sqrt{\sum_{k=1}^{N} \frac{(\widetilde{d_k} - d_k)^2}{N}}, \quad (17)$$

where $\widetilde{d_k}$ are the predicted values, $d_k$ are the corresponding ground truth values, $N$ denotes the total number of measurement points over space and time, and $\bar{d}$ is their mean. NRMSE is frequently used for the comparison between data sets or models with different scales. The smaller the NRMSE, the more accurate the predictions of the model are.

For a fair comparison, (except for the AE+LSTM approach that is limited to 1-step predictions), *the output of all deep learning based approaches evaluated, including the STN, and standard MLP, ConvLSTM, 3D-ConvNet, and SVM, are mixed with the empirical mean $\overline{D}_{t+h}$ of the available ground truth*. That is

$$D'_{t+h} := \gamma(h)D^M_{t+h} + (1 - \gamma(h))\overline{D}_{t+h}. \quad (18)$$

We illustrate the mobile data traffic volume predicted by (D-)STN and conventional forecasting techniques, alongside with ground truth observations in Milan (above) and Trentino (below) in Fig. 5. Each sub-figure shows a different scenario, in terms of the time of day when the predictors are triggered. For the deep learning methods, observations are carried out for 2 hours (data shown before the vertical dotted lines) and predictions are performed for a total duration of 10 hours. Note that the 2h-long observations are used as *input* for prediction and not for actual training. HW-ExpS and ARIMA are continuously fed with ground truth data, as required.

Observe that in predictions in the city of Milan, **STN and D-STN yield the best performance among other approaches, especially as the prediction duration grows.** In particular, traditional HW-ExpS largely over-/under-estimates future traffic, ARIMA gives almost linear and slowly increasing estimates, while the other machine learning based approaches fail to work in the absence of timely observations. HW-ExpS employs a constant weight for its seasonal component, which is independent of the prediction step. Hence, the predictions made may deviate immediately if the data exhibits frequent fluctuation at the initial prediction steps. The predictions made with ARIMA tend to converge to the traffic mean, hence the almost flat long-term behaviour.

Further, we compute the average and standard deviation of the NRMSE attained by all approaches over 11 prediction instances in Milan, 7 of these performed during weekdays and 4 over weekends, with different number of prediction steps employed. For each prediction instance, **most approaches forecast traffic consumption at all locations across the city for over 60 time steps for all instances**. We only employ AE+LSTM for one-step predictions, as this does not support longer forecasting. We summarise these results in Table 1 where we also include the traditional SVM for completeness. Observe that **D-STN performs best in all cases**, attaining superior performance to that of other neural network based schemes (i.e. Conv-LSTM, 3D-ConvNet, and MLP) used individually, **which confirms the effectiveness of combining such elements through multiple fusion layers**, as we propose. In addition D-STN achieves NRMSEs up to 60% and 38% smaller than those obtained with HW-ExpS and respectively ARIMA, and as expected SVM is inferior to the MLP. Our approach further outperforms the AE+LSTM solution in terms of one-step prediction by 26%, which confirms the superior performance of our architecture.

We now turn attention to the performance of all schemes on the Trentino data set. Recall that the deep learning based prediction schemes were trained on the Milan data, whilst HW-ExpS, ARIMA, and AE+LSTM had to be retrained to capture the traffic features of Trentino. Also in this case, we first examine the volume of traffic predicted by our (D-)STN proposals and the other approaches during one weekend day, as shown in the lower sub-plots of Fig. 5. Note that the traffic volume is relatively small (8 Mbytes consumed on average every 10 minutes) and consumption patterns during weekdays are very similar. Observe that the **output of STN follows very closely the real measurements, while the performance of D-STN is nearly identical** (curves overlap). However, the ARIMA predictor does not follow the ground truth, but instead yields a nearly constant output that is close to the average of the measurements. The HW-ExpS substantially underestimates the volume of traffic (except when triggered at 18h).

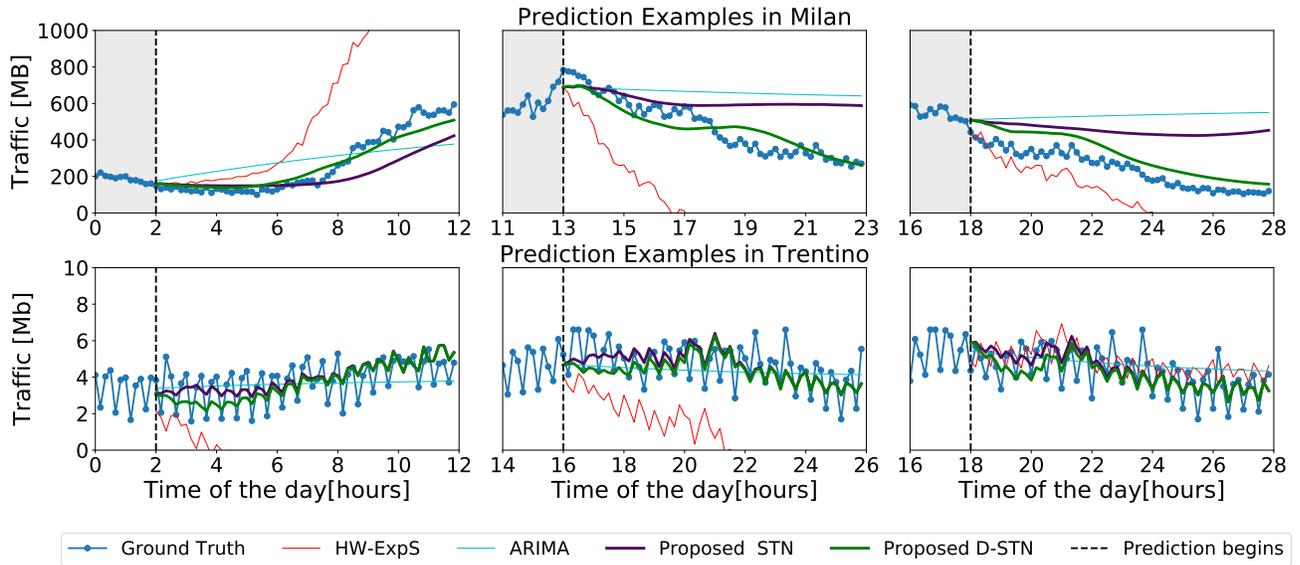

**Figure 5:** Traffic predicted by the proposed (D-)STN and conventional approaches, as well as ground truth measurements in Milan (above) and Trentino (below). Curves in each plot correspond to predictions started at different time of day. Experiments performed at location index (46,57) in Milan during 23$^{rd}$ Dec 2013 and location index (50,62) in Trentino on 21$^{st}$ Dec 2013.

We further compute averages and standard deviation of the mean over 11 prediction instances and examine the performance over prediction durations of different lengths (1 to 60 steps). We summarise these results in Table 2, where we observe that on average ARIMA is slightly more accurate in short-term predictions as the volume of traffic remains low, while **STN performs substantially better in the long-term (61% and 35% lower NRMSE than HW-ExpS and ARIMA)**. We note that since the OTS is employed over the Milan data set, in this case D-STN may be unnecessarily sophisticated, given the light and periodic nature of the traffic. Specifically, each data point corresponds to a measurement in a 235m ×235m cell in Milan, while in Trentino the granularity is relative to 1000m ×1000m squares. Further, data consumption in Milan over 10 minutes is almost 8 times higher than in Trentino. However, the performance of D-STN remains very close to that of the STN in Trentino. Short-term, our proposals yield again lower NRMSEs than AE+LSTM. Nonetheless, as we did not have to retrain (D-)STN with the Trentino data set, the superior performance of our approach also confirms its excellent generalisation abilities.

| Method | 1-step | 10-step | 30-step | 60-step |
|---|---|---|---|---|
| STN | **0.19±0.02** | 0.29±0.05 | 0.51±0.12 | 0.83±0.14 |
| D-STN | **0.19±0.02** | 0.28±0.04 | 0.48±0.09 | 0.71±0.18 |
| HW-ExpS | 0.33±0.03 | 0.51±0.08 | 0.96±0.01 | 1.79±0.45 |
| ARIMA | 0.20±0.04 | 0.39±0.15 | 0.77±0.31 | 1.00±0.27 |
| MLP | 0.23±0.02 | 0.38±0.03 | 0.67±0.13 | 0.96±0.22 |
| ConvLSTM | 0.23±0.02 | 0.39±0.05 | 0.95±0.24 | 1.49±0.24 |
| 3D-ConvNet | 0.20±0.02 | 0.37±0.09 | 0.95±0.30 | 1.64±0.34 |
| SVM | 0.39±0.16 | 0.46±0.11 | 0.62±0.14 | 0.95±0.19 |
| AE+LSTM | 0.24±0.05 | – | – | – |

**Table 1:** NRMSE (mean±std) comparison between different predictors over the Milan data set. Eleven prediction instances triggered at different times of day are used to compute statistics in each case.

| Method | 1-step | 10-step | 30-step | 60-step |
|---|---|---|---|---|
| STN | 0.47±0.03 | 0.66±0.10 | **0.76±0.11** | **0.85±0.07** |
| D-STN | 0.47±0.03 | 0.68±0.11 | 0.78±0.11 | 0.86±0.08 |
| HW-ExpS | 0.57±0.07 | 0.76±0.08 | 1.28±0.16 | 2.19±0.49 |
| ARIMA | **0.38±0.09** | **0.61±0.24** | 0.95±0.34 | 1.30±0.35 |
| MLP | 0.67±0.08 | 0.85±0.05 | 0.85±0.08 | 0.86±0.07 |
| ConvLSTM | 0.52±0.05 | 0.72±0.07 | 0.80±0.10 | 0.89±0.08 |
| 3D-ConvNet | 0.48±0.04 | 0.67±0.07 | 0.77±0.09 | 0.84±0.07 |
| SVM | 0.40±0.04 | 0.55±0.02 | 0.92±0.15 | 1.63±0.55 |
| AE+LSTM | 0.50±0.06 | – | – | – |

**Table 2:** NRMSE (mean±std) comparison between different predictors over the Trentino data set. Eleven prediction instances triggered at different times of day are used to compute statistics in each case.

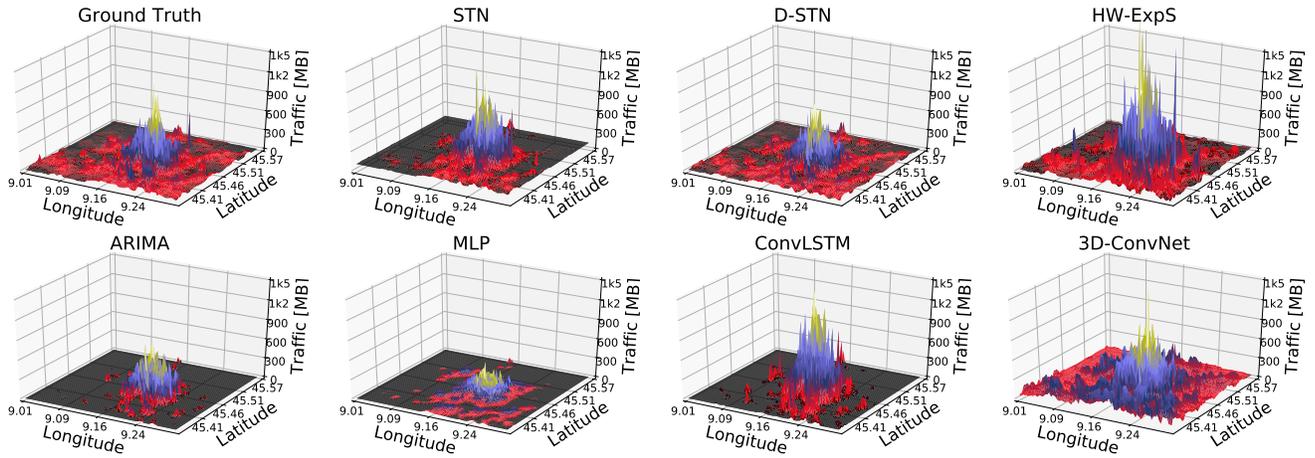

**Figure 6: Snapshots of network-wide predictions made after 10 hours by the proposed (D-)STN, and existing deep learning based and traditional forecasting approaches in Milan, on 24th Dec 2013.**

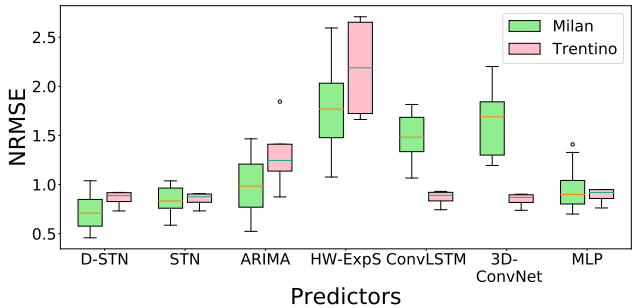

**Figure 7: Long-term (10h) prediction statistics (quartiles of NRMSE) for the proposed (D-)STN, conventional, and other deep learning based approaches in Milan and Trentino.**

To give additional perspective on the value of employing our proposals for long-term mobile traffic forecasting, in Fig. 6 we show snapshots of the predictions made by all approaches considered, across the *entire* network. These are taken at instances that are 10h after measurements have been suspended in the city of Milano. Observe that **D-STN delivers the best prediction among all approaches**. STN works well at city centre level, while slightly underestimating the traffic volume in the surrounding areas. HW-ExpS significantly overestimates the traffic volume, particularly in the central city. The other approaches (ARIMA, MLP, ConvLSTM, and 3D-ConvNet), although successfully capturing the spatial profile of mobile traffic, under-/over-estimate the traffic volume.

We provide further evidence of the superior performance of the proposed (D-) STN methods in forecasting long-term data traffic consumption, by investigating the distribution of the long-term predictions (60 steps) of all methods considered when these are triggered at different times of the day and during both weekdays and weekends. To this end, we show in Fig. 7 with box-and-whisker plots the quartiles of the NRMSE (min, max, median, 1st and 3rd quartiles) for all approaches in both Milano and Trentino areas. Observe that **both STN and D-STN make consistently good predictions, as their median is the lowest in all cases, while the difference between quartiles is small**.

Not least, it is important to note that linear time series prediction methods require longer observations to forecast the traffic volume. In particular the 50-day long **data series used for ARIMA and HW-ExpS are 600 times longer than those employed by the proposed deep learning approaches**. Moreover these techniques cannot be reused without prior re-training in each city. In contrast, **once trained with one data set, (D-)STN can be easily re-used in other cities and provide excellent performance, even if the geographical layouts and traffic volumes differ**.

We conclude that, by exploiting the spatio-temporal features of mobile traffic, the proposed D-STN and STN schemes substantially outperform conventional traffic forecasting techniques (including HW-ExpS and ARIMA) as they attain up to 61% lower prediction errors (NRMSE), do not require to be trained to capture the specifics of a particular city, and operate with up to 600 times shorter measurement intervals, in order to make long-term (up to 10h) predictions. (D-)STN further outperform other deep learning based approaches, such as ConvLSTM, 3D-ConvNet, MLP, and SVM.

## 6 RELATED WORK

**Mobile Traffic Forecasting:** Several time series prediction schemes have been proposed to understand and predict mobile traffic dynamics [20, 27, 33]. Widely used prediction techniques such as Exponential Smoothing [27] and ARIMA [33] employ linear time series regression. Tikunov and Nishumura introduce a Holt-Winters exponential smoothing scheme [31] for short-term forecasting based on historical data in GSM/GPRS networks [27]. Similarly, ARIMA is employed to predict data traffic across 9,000 base stations in Shanghai [33]. These designs estimate user traffic at individual base stations, while ignoring important spatial correlations [20]. They only consider prior temporal information and require continuous long observation periods to perform well. Non-trivial spatio-temporal patterns of mobile traffic were recently analysed by Exploratory Factor Analysis (EFA) [8] for the purpose of network activity profiling and land use detection.

**Deep Learning Based Predictors:** Recent deep learning advances substantially improved the performance of image recognition and natural language processing [5, 18]. Recurrent Neural Networks (RNNs) such as LSTM [9] outperform traditional machine learning approaches in terms of time series prediction accuracy. An advanced version of RNNs called ConvLSTM has been applied for precipitation nowcasting [32]. 3D-ConvNets show remarkable performance in terms of spatio-temporal feature learning in video and 3D imaging [12, 13], while they differ from RNNs as they do not train by back propagation through time (BPTT). Their time series forecasting abilities remain largely unexplored. Recently, Wang et al. employ local and global auto-encoders to extract spatial features from mobile traffic, and subsequently use an LSTM to perform predictions [30]. However, their approach *requires to train an auto-encoder for each individual location*, which is computationally inefficient for large-scale forecasting tasks. Moreover, this approach can only perform very short-term (i.e. one step) forecasting, which limits its applicability to real deployments.

Our work exploits the potential of deep learning (i.e. ConvLSTM and 3D-ConvNet) to perform *long-term* mobile traffic volume predictions. The results we obtain demonstrate the proposed (D-)STN schemes can predict mobile traffic over up to 10 hours with excellent accuracy, while working with limited observations (2 hours).

## 7 CONCLUSIONS

Forecasting traffic in cellular networks is becoming increasingly important to dynamically allocate network resources and timely respond to exponentially growing user demand. This task becomes particularly difficult as network deployments densify and the cost of accurate monitoring steepens. In this paper, we proposed a Spatio-Temporal neural Network (STN), which is a precise traffic forecasting architecture that leverages recent advances in deep neural networks and overcomes the limitations of prior forecasting techniques. We introduced an Ouroboros Training Scheme (OTS) to fine tune the pre-trained model. Subsequently, we proposed a Double STN (D-STN), which employs a light-weight mechanism for combining the STN output with historical statistics, thereby improving long-term prediction performance. Experiments conducted with publicly available 60-day long traffic measurements collected in the city of Milan and the Trentino region demonstrate the proposed (D-)STN provide up to 61% lower prediction errors as compared to widely employed ARIMA and HW-ExpS methods, while requiring up to 600 times shorter ground truth measurement durations.